# Speckle-interferometric study of close visual binary system Hip 11253 (HD14874) using Gaia (DR2) and (EDR3)

Hussam Aljboor[1] and Ali Taani[2]

[1] Department of Basic Scientific Sciences, Prince Hussein Bin Abdullah II Academy for Civil Protection, Al Balqa Applied University, Amman, Jordan
Email: Hussamaljboor32@gmail.com

[2] Physics Department, Faculty of Science, Al Balqa Applied University, 19117 Salt, Jordan
Email: ali.taani@bau.edu.jo

## ABSTRACT

We present a comprehensive set of physical and geometrical parameters for each of the components of the close visual binary system Hip 11253 (HD14874). We present an analysis for the binary and multiple stellar systems with the aim to obtain a match between the overall observational spectral energy distribution of the system and the spectral synthesis created from model atmospheres using Al-Wardat's method for analyzing binary and multiple stellar systems. The epoch positions are used to determine the orbital parameters and the total mass. The parameters of both components are derived as: $T_{eff}^a = 6025, T_{eff}^b = 4710, \log g_a = 4.55, \log g_b = 4.60, R_a = 1.125 R_\odot, R_b = 0.88 R_\odot, L_a = 1.849 L_\odot, L_b = 0.342 L_\odot$. Our analysis shows that the spectral types of both are F9a and K3b. By combining the orbital solution with the parallax measurements of Gaia DR2 and EDR3, we estimate the individual masses using H-R diagram $M_a = 1.09 M_\odot$ and $M_b = 0.59 M_\odot$ for using Gaia DR2 parallax and $M_a = 1.10 M_\odot$ and $M_b = 0.61 M_\odot$ for using Gaia EDR3 parallax. Finally, the location of both system's components on the stellar evolutionary tracks is presented.

# 1. INTRODUCTION

About half of the stars in the galaxy are binaries or stellar systems (Duquennoy & Mayor, 1991; Cai *et al.*, 2012). The analysis of binary systems is an efficient method for determining the physical and geometrical properties, such as the stellar masses. Close Visual Binary Stars (CVBSs) are binaries with a small separation angle (1 arcsecond or less) between their components, and they are hard to detect with a small telescope. Since the long orbital periods of CVBSs have a range of 10-1000 yrs., it takes a long time to observe one orbital period (Jiang *et al.*, 2013; Taani, Abushattal and Mardini 2019). The visual binaries are an important source to obtain the masses and distances of the stars, which contribute to the understanding of their basic physical properties (Taani & Vallejo 2017; Taani, Vallejo and Abu-Saleem 2022; Taani *et al.*, 2022). The recent measurements of such systems by the Gaia astrometry mission with increased precision have improved the distances to such systems (Collaboration 2018; Mardini et al. 2019a,b; Mardini et al. 2020; Brown et al. 2021). The system HIP11253 being visually close enough, enable accurate measurements of its colors, color indices, and magnitude difference between its components.

In this paper, we use two methods. The first one is Al-Wardat's method for analyzing binary and multiple stellar systems (BMSSs), which was introduced by (Al-Wardat 2002), and used to analyze many CVBSs and BMSSs (Al-Wardat and Widyan, 2009; Al-Wardat, 2012; Masda et al., 2019; Widyan and Aljboor 2021; Tawalbeh et al., 2021; Al-Wardat et al., 2021; Abu-Dhaim et al. 2022). The main idea of Al-Wardat's method is to build entire spectral energy distributions (SEDs) of BMSSs using the model atmospheres like those of (ATALS9) (Kurucz 1994), by implementing the available observational measurements like magnitude difference and colour indices.. The second method is Tokovinin's method for the dynamical analysis of binary stars (Tokovinin 1992), which was used to solve the orbit of the system. Details of this program can be found by (Tokovinin 2017).

HIP11253 is located at a right ascension of $02^h 20^m 51^s$, and declination of $+30°38'48''$ (SIMBAD catalog). The parallax of the system is obtained from the Gaia Data Release 2 (Collaboration 2018) and it's a value is $18.9878 \pm 0.627$ mas, which corresponds to a distance of 52.67pc and the Gaia Early Data Release 3 (Brown et al., 2021) and its value is $18.1854 \pm 0.213$ mas, which corresponds to a distance of 54.99pc. Table (1), contains the basic information on the system from the (SIMBAD) and other databases.

The aim of this study is to find reliable stellar parameters for systems that may produce the best agreement between measured magnitudes and color indices and synthetic results. In addition, we use the Gaia parallaxes DR2 (Collaboration 2018) and Gaia parallaxes EDR3 (Brown et al., 2021). We emphasize that the binary system HIP11253 is presented as an example of how to use the approach mentioned above in the case of a visual close binary of low total mass.

| Properties | Parameters | Value | *Reference* |
|---|---|---|---|
| Position | $\alpha_{2000}$ | $02^h 20^m 51^s$ | (*SIMBAD Astronomical Database - CDS (Strasbourg)*, no date) |
| | $\delta_{2000}$ | $+30°38'48''$ | (*SIMBAD Astronomical Database - CDS (Strasbourg)*, no date) |
| Magnitudes [mag] | $m_v$ | 8.16 | (ESA, 1997) |
| | $A_v$ | 0.2432 | (Schlafly and Finkbeiner, 2011) |
| | $(B-V)_J$ | $0.665 \pm 0.018$ | (ESA, 1997) |
| | $B_T$ | $8.96 \pm 0.012$ | (Hog *et al.*, 2000) |
| | $V_T$ | $8.23 \pm 0.010$ | (Hog *et al.*, 2000) |
| Parallax [mas] | $\pi_{GDR2}$ | $18.9878 \pm 0.627$ | (Brown *et al.*, 2018) |
| | $\pi_{GEDR3}$ | $18.1854 \pm 0.213$ | (Brown *et al.*, 2021) |

**Table 1: Basic parameters and observed astrometric and photometric data of HIP 11253.**

| $\Delta m$ | $Tel^\star$ | $filter(\lambda/\Delta\lambda)$ | Reference |
|---|---|---|---|
| $1.89 \pm 0.03$ | 6.0 | $545nm/30$ | (Pluzhnik, 2005) |
| $2.79 \pm 0.22$ | 6.0 | $545nm/30$ | (Balega *et al.*, 2002) |
| $2.48 \pm 0.19$ | 6.0 | $600nm/30$ | (Balega, Balega and Maksimov, 2006) |
| $2.95 \pm 0.00$ | 3.5 | $550nm/40$ | (Horch *et al.*, 2008) |
| $2.68 \pm 0.05$ | 6.0 | $540nm/30$ | (Balega *et al.*, 2007) |
| $2.71 \pm 0.00$ | 3.5 | $550nm/30$ | (Horch *et al.*, 2009) |
| $2.55 \pm 0.00$ | 3.5 | $562nm/40$ | (Roberts, 2011) |

**Table 1: Magnitude difference between the components of the system Hip11253, along with the filters used to obtain the obsrvations.**

| Epoch | $\theta$ | $\delta\theta$ | $\rho$ | $\delta\rho$ | mcth | Reference |
|---|---|---|---|---|---|---|
| 1991.25 | 362 | . | 0.239 | . | Hh | (Hartkopf *et al.*, 1997) |
| 1998.7747 | 285.6 | 0.6 | 0.344 | 0.004 | S | (Balega *et al.*, 2002) |
| 1998.9246 | 285.8 | . | 0.347 | . | S | (Horch *et al.*, 2002) |
| 1999.8130 | 283.9 | 0.3 | 0.351 | 0.002 | S | (Balega *et al.*, 2004) |
| 2000.8730 | 282.7 | 0.5 | 0.356 | 0.003 | S | (Balega *et al.*, 2006) |
| 2001.7528 | 282.1 | 0.4 | 0.362 | 0.002 | S | (Balega *et al.*, 2006) |
| 2002.7992 | 278.2 | 1.1 | 0.386 | 0.007 | S | (Balega *et al.*, 2013) |
| 2003.6290 | 277.8 | . | 0.372 | . | S | (Horch *et al.*, 2008) |
| 2003.6290 | 275.7 | . | 0.368 | . | S | (Horch *et al.*, 2008) |
| 2004.8239 | 276.9 | 0.3 | 0.372 | 0.002 | S | (Balega *et al.*, 2007) |

| 2007.8256 | 273.2 | .    | 0.379 | .     | S  | (Tokovinin, Mason and Hartkopf, 2010) |
| 2009.7371 | 279.7 | 21.7 | 0.350 | 0.05  | S  | (Voitsekhovich and Orlov, 2014) |
| 2010.0100 | 269.5 | .    | 0.373 | .     | S  | (Horch *et al.*, 2011) |
| 2012.6777 | 265.7 | 0.3  | 0.366 | 0.009 | S  | (Riddle *et al.*, 2015) |
| 2013.7976 | 264.1 | .    | 0.344 | .     | Ag | (Kehrli *et al.*, 2017) |

**Table 2:** New data of interferometric Measurments for the HIP 11253 system.

| Orbital element | Unit | Last work (Ling 2011) | This work |
|---|---|---|---|
| $P$ | $[yr]$ | 82.18 | $84.48 \pm 0.86$ |
| $\omega$ | $[deg]$ | 278.8 | $276.11 \pm 0.27$ |
| $e$ | $--$ | 0.283 | $0.288 \pm 0.0054$ |
| $\Omega$ | $[deg]$ | 87.95 | $86.95 \pm 0.16$ |
| $i$ | $[deg]$ | 109.7 | $108.61 \pm 0.25$ |
| $T$ | $[yr]$ | 2026.35 | $2026.80 \pm 0.2184$ |
| $a$ | $[arcsec]$ | 0.378 | $0.3921 \pm 0.008$ |
| $M_T^*$ | $[M_\odot]$ | 1.1682 | 1.2342 |
| $M_T^{**}$ | $[M_\odot]$ | 1.3297 | 1.4049 |
| $RMS(\theta)$ | $[deg]$ | 0.62 | 0.5907 |
| $RMS(\rho)$ | $[arcsec]$ | 0.0124 | 0.0055 |

**Table 3:** Orbital parameters solutions and total masses formerly published for the HIP 11253 system, for comparison with this work. *Using parallax Gaia Data Release 2 (DR2). ** Using parallax Gaia Data Release 3 (EDR3).

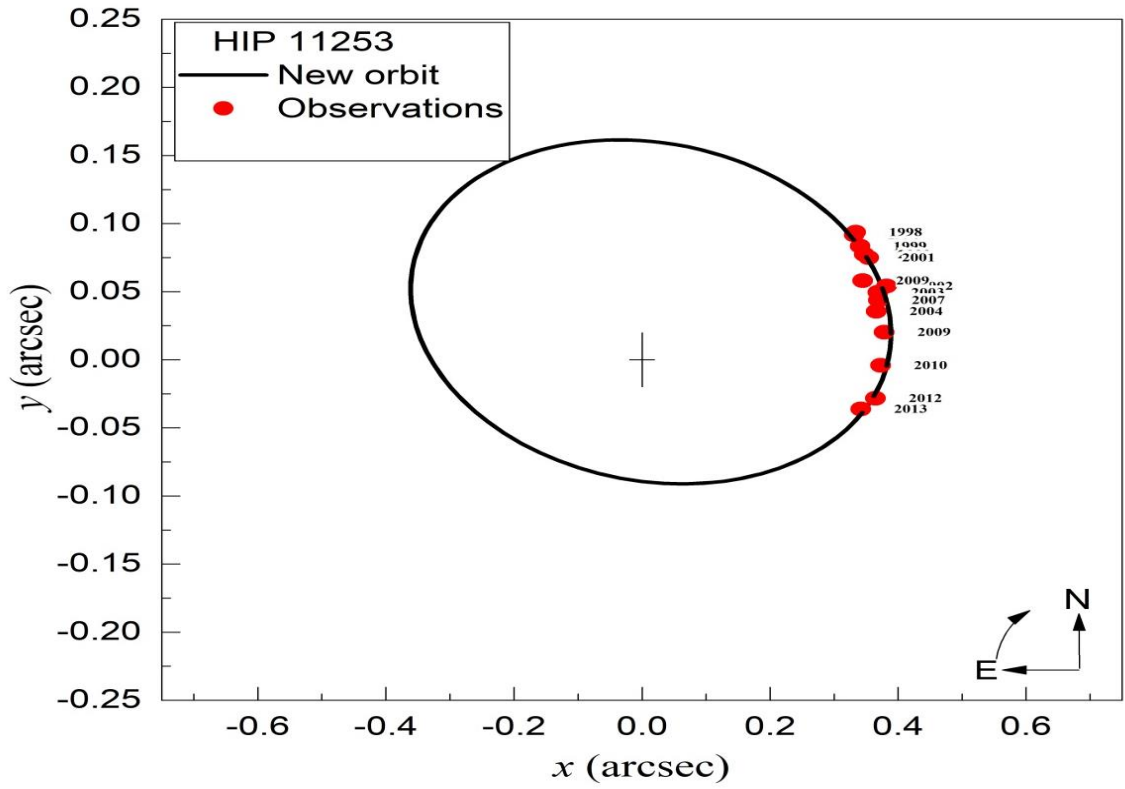

**Figure 1:** The relative orbit of the binary system HIP 11253 constructed using the relative position. The measurements were taken from the fourth catalog of Interferometric measurements of binary stars.

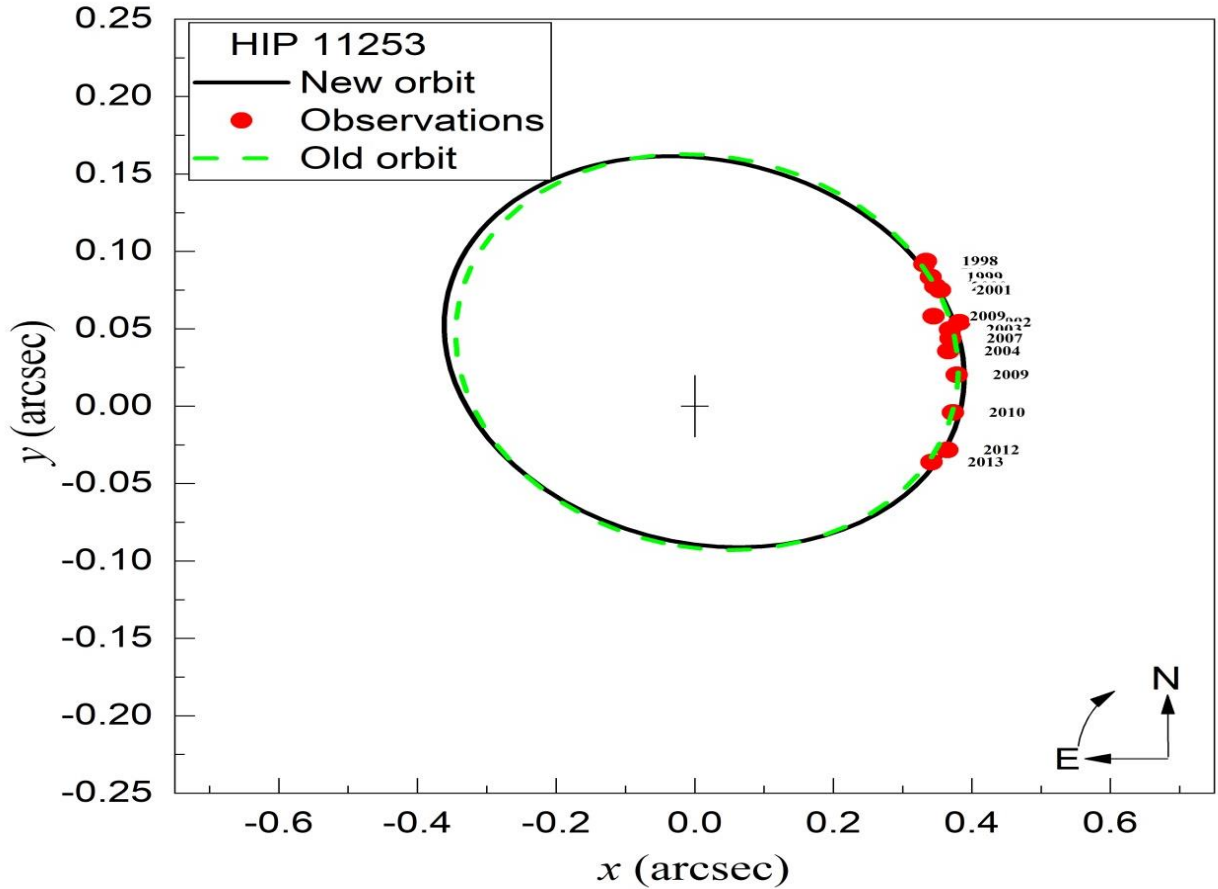

**Figure 2: The different in the orbit between the last work from Ling (2011) and this work.**

## 2. ORBTIAL ELEMENTS:

The determination of the orbital elements is crucial in calculating the stellar masses. The orbit of the HIP11253 system is obtained using the position angles ($\theta$) and the angular separations ($\rho$) obtained from the fourth Catalog of Interferometric Measurement of binary stars (INT4) (see table 3), following Tokovinin's method (Tokovinin 1992). The orbit is shown in figure (1), and the results of the dynamical analysis and orbital solutions of HIP 11253 are listed in Table (4).

Table (4) shows a good agreement between our estimated orbital period (P); inclination (i); semi-major axis (a); eccentricity (e); position angle of nodes ($\Omega$); argument of periastron ($\omega$); and time of primary minimum (T) with those previously reported results.

Figure (2) depicts the new orbit in comparison to the old one obtained earlier by (Ling 2011). We have estimated the total dynamical mass for this binary system, using the Kepler's third law. The total dynamical mass is given by:

$$M_{Dyn} = M_A + M_B = \left(\frac{a}{\pi}\right)^3 \frac{M_\odot}{P^2}, \qquad (2.1)$$

where a is the semi-major axis in (arcsec), π is the trigonometric parallax in (arcsec), $M_A, M_B$ are the masses of the individual components, $M_{dyn}$ is the dynamical mass sum, P: the relative orbital period, and $M_\odot$ is the mass of the Sun equal to $1.9891 \times 10^{30}$ kg.

The error of the dynamical mass sum is given by:

$$\frac{\delta_M}{M_{Dyn}} = \sqrt{9\left(\frac{\delta_\pi}{\pi}\right)^2 + 9\left(\frac{\delta_a}{a}\right)^2 + 4\left(\frac{\delta_P}{P}\right)^2} \qquad (2.2)$$

Using the orbital period and semi-major axis obtained from the orbital solution and parallax measurement, we estimate the dynamical mass as it appears in table (4).

## 3. ATMOSPHERIC MODELING

### 3.1 INPUT PARAMETERS

We derive the physical parameters of each component of the binary star system HIP11253 by following Al-Wardat's method for analyzing BMSSs (Al-Wardat 2002). This is done by creating individual SED models for each system component using the visual magnitude $m_v = 8^m.16$ from Table (1), and the average value for visual magnitude band $\Delta m_v = 2^m.16$ from the speckle interferometric results in see Table (2), along with parallax of the system ($\pi = 18.9878, d = 52.67 pc$) from Gaia DR2, ($\pi = 18.1854, d = 54.99 pc$) from Gaia EDR3, and the bolometric corrections (BC) by Lang(Lang 1992). The preliminary specific value of the components input parameters were estimated as seen in Table (5). To determine the apparent visual magnitude of individual components ($m_v^A$ and $m_v^B$) we use the following equations:

$$m_v^A = m_v + 2.5 \log(1 + 10^{-0.4\Delta m_v}), \qquad (3.1)$$
$$m_v^B = m_v^A + \Delta m_v \qquad (3.2)$$

$$M_v = m_v + 5 + 5 \log \pi - A_v \qquad (3.3)$$

Where $A_v$ is the interstellar extinction coefficient, $\pi$ the trigonometric parallax (Brown *et al.*, 2018). The absolute magnitude of individual components ($M_v^A$ and $M_v^B$) can be calculated according to the equations in Lang (1992) and Gray (2005). With the absolute magnitude, the effective temperature ($T_{eff}$), and the BC can be obtained. The absolute bolometric magnitude ($M_{bol}$) is then calculated using BC.

$$M_{bol} = M_V - BC \qquad (3.4)$$

$$M_{bol}^* - M_{bol}^\odot = -2.5 \log\left(\frac{L_*}{L_\odot}\right). \qquad (3.5)$$

Where: $M_{bol}^\odot$: is the bolometric magnitude of the Sun= $4^m.75$, $L_\odot$ is Luminosity of the Sun $= 3.79 \times 10^{26}$ Watt, $R_\odot$ is the radius of Sun $= 6.957 \times 10^5$ km, $T_\odot$ is the effective temperature of Sun $= 5777$ K, and $M_\odot$ is the Mass of Sun $1.988 \times 10^{30}$ kg.

The radius (R) is obtained through:

$$\log\left(\frac{R}{R_\odot}\right) = 0.5 \log\left(\frac{L}{L_\odot}\right) - 2 \log\left(\frac{T}{T_\odot}\right). \qquad (3.6)$$

The gravitational acceleration ($\log g$) is obtained as:

$$\log g = \log\left(\frac{M}{M_\odot}\right) - 2 \log\left(\frac{R}{R_\odot}\right) + 4.43. \qquad (3.7)$$

Note that the $T_{eff}$ and $\log g$ values for both components are considered as the preliminary input parameters for both components' atmospheric modeling. As a result, we can compute their synthetic spectra.

| Parameters | Unit | Gaia parallax **DR2** [(Brown *et al.*, 2018) | | Gaia parallax **EDR3** (Brown *et al.*, 2021) | |
|---|---|---|---|---|---|
| | | A | B | A | B |
| $m_V$ | [mag] | 8.30 | 10.46 | 8.30 | 10.64 |
| $M_V$ | [mag] | 4.45 | 6.61 | 4.36 | 6.52 |
| BC* | [mag] | $-0.09$ | $-0.41$ | $-0..08$ | $-0.41$ |
| $M_{bol}$ | [mag] | 4.61 | 7.05 | 4.44 | 6.93 |
| $T_{eff}$* | [k] | 5987 | 5080 | 6030 | 4815 |
| R | [$R_\odot$] | 2.976 | 1.342 | 3.172 | 1.578 |
| L | [$L_\odot$] | 1.138 | 0.120 | 1.330 | 0.134 |
| Sp − Type* | − − | G0 | K2 | F9 | K3 |

**Table 4: Preliminary physical and geometrical properties of HIP 11253. The properties marked by asterisks are obtained using the tables from Lang (1992) and Gray (2005).**

## 3.2. Synthetic Spectra:

The input parameters obtained in sec. 3.1 are adopted to model the single star's atmospheres by using the blanked model (ATLAS 9) due to Kurucz's (Kurucz 1994). While the entire SEDs were built using special subroutines of Al-Wardat's method. Another subroutine of the same method was also used to calculate the individual and entire SEDs. The following equation is used to calculate the synthetic SED as observed from the Earth, which is related to the energy flux of each component (Al-Wardat 2002):

$$F_\lambda . d^2 = H_\lambda^A . R_A^2 + H_\lambda^B . R_B^2 \qquad (3.8)$$

which can be expressed as:

$$F_\lambda = (R_A/d)^2 [H_\lambda^A + H_\lambda^B \cdot (R_B/R_A)^2] \qquad (3.9)$$

Where, $H_\lambda^A, H_\lambda^B$ are the fluxes at the surface of the star, $F_\lambda$ is the flux for the entire (SED) of the binary system, and $R_A, R_B$ are the radii of the primary and secondary components of the system in solar units.

Several works applied an iterative scheme using a different combination of input parameters to get the best fit between the observed flux and the total computed flux (see i.e, Al-Wardat and Widyan, 2009). Additionally, multiple values of $\Delta m$, $m_v$, and parallax to achieve convergence. In this way, various models are obtained and compared with the observed SED. The results are presented in the figures (3 \& 4) and in Table (6).

Al-Wardat's method criteria to find the best fit are based on the following: the inclination of the spectra, maximum values of the fluxes, and profiles of the absorption lines. As a consequence, the best fit we have found (see figures 3 and 4) was found using the following set of parameters (see Table 6).

However, Table (7) shows the synthetic spectrum for the entire and individual components of HIP 11253. While the comparison between the observational and synthetic values for the color index and magnitude differences is listed in Table (8). A good indication of the reliability of the obtained parameters of the different system components as listed in Table (9). In addition, based on the ultimate effective temperatures of the system, we estimate the bolometric magnitudes and the stellar luminosities, by using the following equation :

$$\log\left(\frac{R}{R_\odot}\right) = 0.5 \log\left(\frac{L}{L_\odot}\right) - 2 \log\left(\frac{T}{T_\odot}\right). \qquad (3.10)$$

$$\log g = \log\left(\frac{M}{M_\odot}\right) - 2 \log\left(\frac{R}{R_\odot}\right) + 4.43. \qquad (3.11)$$

The best fit values are sufficiently representative of the system component parameters. Using Lang (1992) and Gray (2005) empirical relations of spectral types and effective temperature, the spectral types of HIP 11253 components can be obtained as ($F9$) and ($K3$) for components a and b respectively.

| HIP 11253 Using | | $\pi_{GDR2}$ | $\pi_{EDR3}$ |
|---|---|---|---|
| **Symbol** | **Units** | | |
| $\pi$ | $[arcsec]$ | 18.9878 | 18.1854 |
| $d$ | $[pc]$ | 52.67 | 54.99 |
| $R_A$ | $[R_\odot]$ | $1.125 R_\odot$ | $1.175 R_\odot$ |
| $R_B$ | $[R_\odot]$ | $0.88 R_\odot$ | $0.92 R_\odot$ |
| $T_{eff}^A$ | $[K]$ | $6025 K$ | $6025 K$ |
| $T_{eff}^B$ | $[K]$ | $4710 K$ | $4710 K$ |
| $\log g_A$ | $[cm/s^2]$ | 4.55 | 4.55 |
| $\log g_B$ | $[cm/s^2]$ | 4.60 | 4.60 |

**Table 6: The final results of this individual components and entire SEDs of the system HIP11253.**

## 4. SYNTHETIC PHOTOMETRY

Following the treatment by Al-Wardat, (2002), the synthetic magnitudes are calculated from the synthetic SED using the following relationship:

$$m_p[F_{\lambda.s}(\lambda)] = -2.5 \log \frac{\int P_p(\lambda) F_{\lambda.s} \lambda d\lambda}{\int P_p(\lambda) F_{\lambda.r} \lambda d\lambda} + ZP_p, \quad (4.1)$$

where $m_p$: is the synthetic magnitude of the passband p, $P_p(\lambda)$: is the dimensionless sensitivity function of the passband p, $F_{\lambda.s}(\lambda)$: is the synthetic SED of the object, $F_{\lambda.r}(\lambda)$: is the SED of reference star Vega, $ZP_p$: the zero-point taken from Maíz Apellániz (2007). By using the iteration method with different sets of observed stellar parameters like magnitude differences of the components and color indices of the entire system and individual components in (U-B-V-R) Johnson-Cousins, (uvby) Strömgren and (B-V) Tycho. One can get the best fit between the observational and synthetic magnitudes. This would help us to judge the accuracy of the estimated parameters using special subroutine of Al-Wardat's method under the Interactive Data Language (IDL) platform and verify the estimated parameter reliability. As we can see, figures (3) and (4) show best-fitting between the individual components and entire synthetic and observational SED. While on other hand, Figures 5 and 6, show the evolutionary tracks of in a theoretical HR diagram of the components of both HIP11253, which are evolving off the main sequence. This helped to estimate the masses and spectral types as in Table (9). Al-Wardat's method for analyzing BMSSs gives a mass sum of $1.68M_\odot$ for parallax Gaia DR2 and $1.71M_\odot$ for parallax Gaia EDR3. It is worth noting that masses estimated using this method along with the modified orbital elements is independent of the parallax value. From the comparison of dynamical masses using this approach, we realize that parallax measurement from EDR3 provides a good indication of the mass sum, due to the advance and higher resolution.

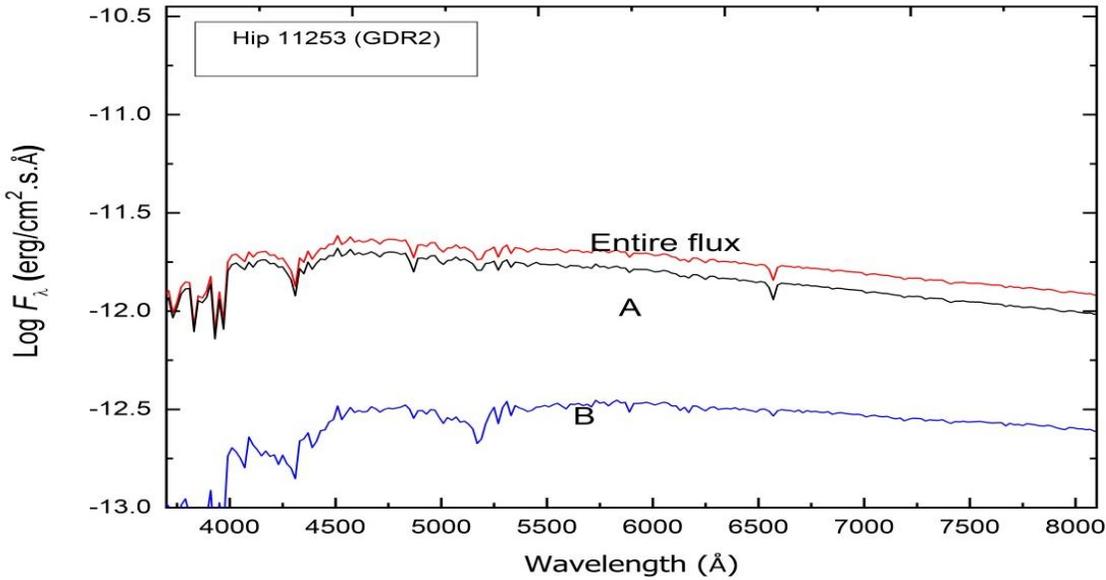

**Figure3: The entire (sum of the fluxes of the members A and B) synthetic SED of the system HIP 11253 (built using parameters of the primary component, $T_{eff} = 6025k, \log g = 4.55 \text{cm/s}^2, R = 1.125 R_\odot$ and the computed flux of the secondary component with $T_{eff} = 4710k, \log g = 4.6055 \text{cm/s}^2, R = 0.88 R_\odot \text{ and } d = 54.99 \text{ pc}$) against the observational one, the figure also shows the synthetic SED for each component**

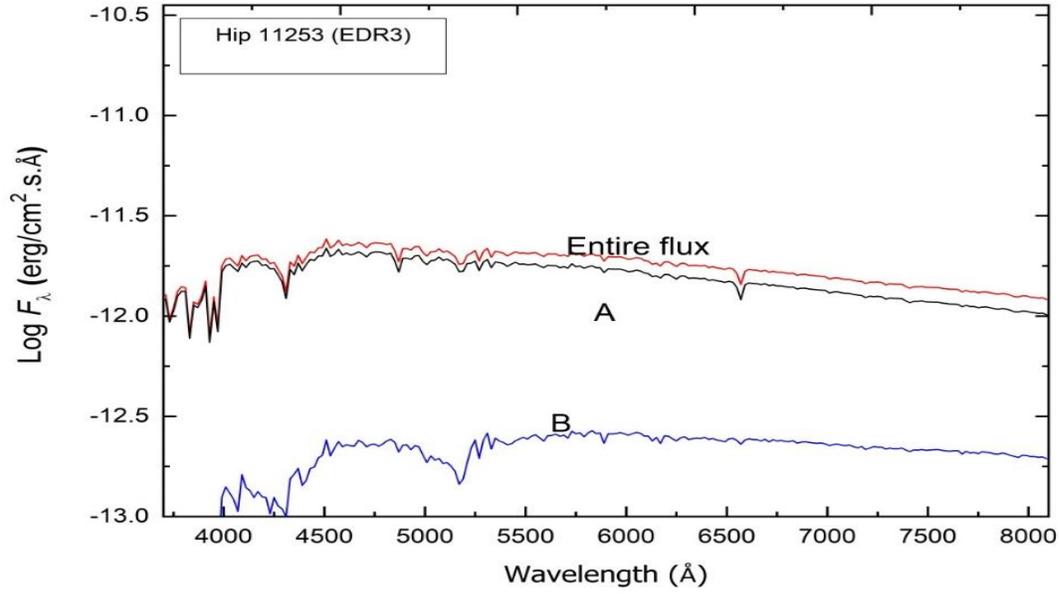

**Figure 4:** The entire synthetic SED of the system HIP 11253 (built using parameters of the primary component, $T_{eff} = 6025k, \log g = 4.55 cm/s^2, R = 1.175 R_\odot$ and the computed flux of the secondary component with $T_{eff} = 4710k, \log g = 4.60 \frac{cm}{s^2}, R = 092 R_\odot$ and $d = 52.67\ pc$) against the observational one, the figure also shows the synthetic SED for each component.

| Sys | Filter | HIP11253 using | | | | | |
|---|---|---|---|---|---|---|---|
| | | Gaia DR2 | | | Gaia EDR3 | | |
| | | Entire Synth. | A | B | Entire Synth. | A | B |
| Joh-Cou. | $U$ | 8.9891 | 9.03494 | 12.4473 | 8.98816 | 9.03410 | 12.4444 |
| | $B$ | 8.82499 | 8.91813 | 11.5380 | 8.82400 | 8.91726 | 11.5351 |
| | $V$ | <u>8.16179</u> | 8.30102 | 10.4605 | <u>8.16070</u> | 8.30023 | 10.4576 |
| | $R$ | 7.78665 | 7.96659 | 9.82692 | 7.78549 | 7.96576 | 9.82400 |
| | $U-B$ | 0.164123 | 0.116807 | 0.99328 | 0.164162 | 0.116833 | 0.909332 |
| | $B-V$ | <u>0.663198</u> | 0.617116 | 1.07748 | <u>0.663301</u> | 0.617037 | 1.07747 |

|  | $V-R$ | 0.375146 | 0.334431 | 0.633608 | 0.375205 | 0.334466 | 0.633633 |
|---|---|---|---|---|---|---|---|
| **Strom.** | $u$ | 10.1379 | 10.1790 | 13.7147 | 10.1369 | 10.1781 | 13.7118 |
|  | $v$ | 9.18100 | 9.25202 | 12.1776 | 9.17996 | 9.25111 | 12.1748 |
|  | $b$ | 8.53023 | 8.64619 | 11.0157 | 8.52918 | 8.64538 | 11.0128 |
|  | $y$ | 8.12762 | 8.27197 | 10.3894 | 8.12655 | 8.27124 | 10.3864 |
|  | $u-v$ | 0.956917 | 0.926981 | 1.53713 | 0.956951 | 0.926999 | 1.53700 |
|  | $v-b$ | 0.650767 | 0.605830 | 1.16186 | 0.650776 | 0.605732 | 1.16202 |
|  | $b-y$ | 0.402607 | 0.374220 | 0.626360 | 0.402628 | 0.374146 | 0.626321 |
| **Tycho.** | $B_T$ | <u>8.98668</u> | 9.06898 | 11.8287 | <u>8.98571</u> | 9.06811 | 11.8258 |
|  | $V_T$ | <u>8.23819</u> | 8.37034 | 10.5902 | <u>8.23711</u> | 8.36952 | 10.5873 |
|  | $B_T - V_T$ | 0.748486 | 0.698636 | 1.23847 | 0.748608 | 0.698564 | 1.23848 |

**Table 5: Magnitudes and color indices of the entire synthetic spectrum and individual components of HIP 11253.**

| | **HIP11253** | | |
|---|---|---|---|
| Filter | Observed[mag] | Synthetic (This work using Gaia DR2) [mag] | Synthetic (This work using Gaia EDR3) [mag] |
| $m_v$ | 8.16 | 8.16179 | 8.1607 |
| $\Delta m$ | 2.16 | 2.15948 | 2.15737 |
| $B_T$ | $8.96 \pm 0.012$ | 8.98668 | 8.98571 |
| $V_T$ | $8.23 \pm 0.010$ | 8.23819 | 8.23711 |
| $(B-V)_J$ | $0.665 \pm 0.018$ | 0.663198 | 0.663301 |

**Table 6: Comparison between the observational and synthetic magnitudes and colours indices for both systems. $m_v$: visual magnitude of the binary system hip 11253 , $\Delta m$: the speckle interferometric results in V-band , $B_T$: Photometric magnitude in Optical B band between 400 and 500 nm Tycho Magnitude, $V_T$: Photometric magnitude in optical V band between 500 and 600 nm Tycho Magnitude, $(B-V)_J$: Color index or magnitude difference between optical B band between 400 and 500 nm and Optical V band between 500 and 600 nm Johnson color index (Synthetic magnitudes were calculated using subtrouties of Al-Wardat's method for analyzing BMSSs).**

| Component | Using Gaia DR2 | | Using Gaia EDR3 | |
|---|---|---|---|---|
| | A | B | A | B |
| $T_{eff}(K)$ | 6025 | 4710 | 6025 | 4710 |
| $\log g$ | 4.55 | 4.60 | 4.55 | 4.60 |
| $Radius(R_\odot)$ | 1.125 | 0.88 | 1.175 | 0.92 |
| $L(L_\odot)$ | 1.8486 | 0.3422 | 1.6334 | 0.3740 |
| $M_{bol}$ | 4.08 | 5.91 | 4.22 | 5.82 |
| $M_V^*$ | 4.16 | 6.39 | 4.30 | 6.30 |
| $Sp.Type^*$ | F9 | K3 | F9 | K3 |
| $parallax[mas]$ | $18.9878 \pm 0.627$ | | $18.1854 \pm 0.213$ | |
| $Mass, (M_\odot)^{**}$ | $M_a = 1.09 M_\odot$ $M_b = 0.59 M_\odot$ $M_a + M_b = 1.68 M_\odot$ | | $M_a = 1.10 M_\odot$ $M_b = 0.61 M_\odot$ $M_a + M_b = 1.71 M_\odot$ | |

**Table 7:** Estimated physical parameters of individual components for HIP 11253 system based on two parallax measurments. The properties marked by asterisks are obtained using the tables from Lang (1992) and Gray (2005). While the double asterisks are obtained using the tables from Girardi et al. (2000). Individual masses were estimated using the results of applying Al-Wardat's method for analyzing BMSSs on the system.

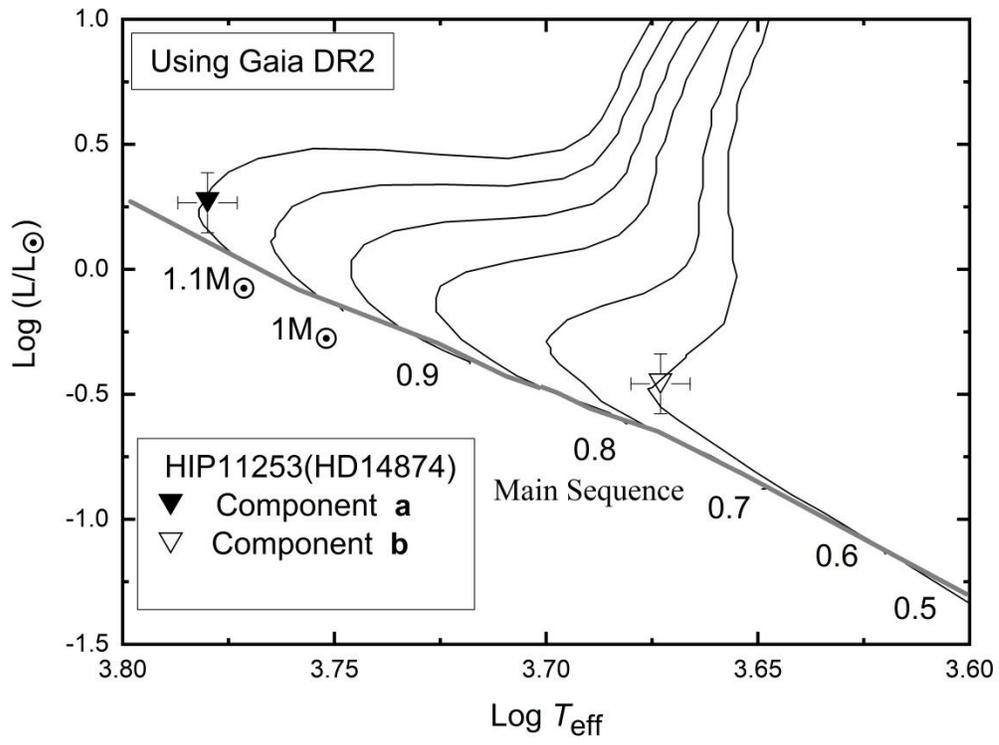

**Figure 5:** The evolutionary tracks of both components of HIP 11253 on the H-R diagram of masses using parallax Gaia DR2. The evolutionart tracks were taken from Girardi et al. (2000).

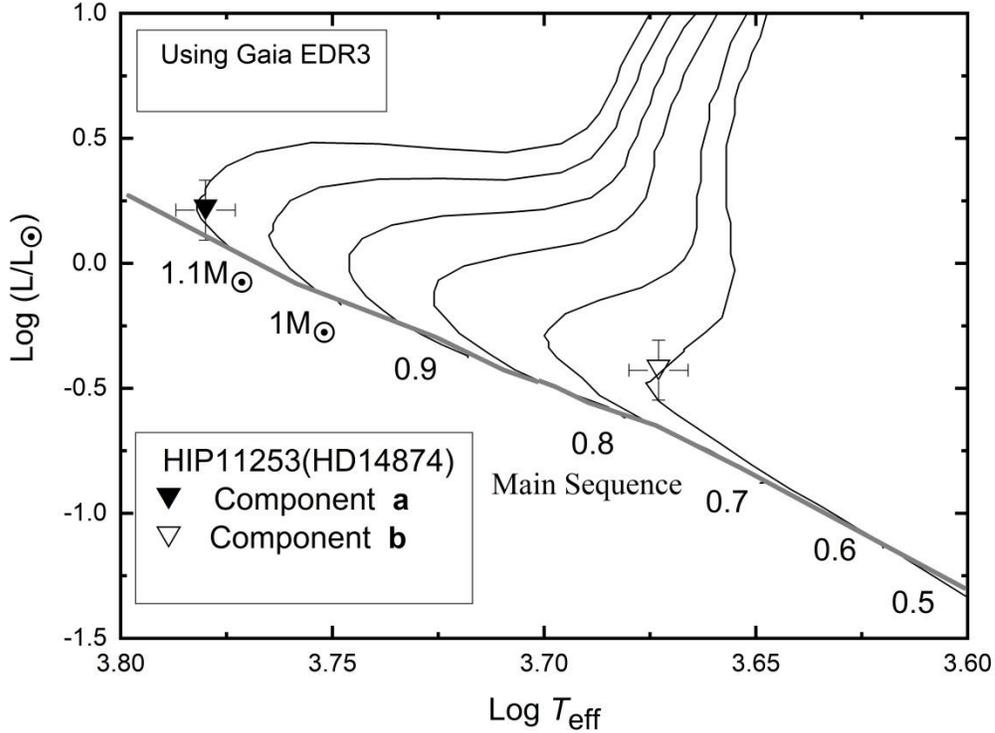

**Figure 6:** The evolutionary tracks of both components of HIP 11253 on the H-R diagram of masses using parallax Gaia EDR3 .The evolutionart tracks were taken from Girardi et al. (2000).

## 5. CONCLUSIONS:

In this work, we have investigated the basic stellar parameters of the close visual binary system Hip 11253 (HD14874). Our investigation was based on the following ingredients:
(a) Creating individual spectral energy distribution (SED) models for each system component as described in sect. (3.1) using Al-Wardat's method for analyzing BMSSs (Al-Wardat 2002), combined with the parallax measurements by the Gaia (DR2 and EDR3).
(b) The orbital elements were calculated following the dynamical method used by Tokovinin (1992).
(c) We used Al-Wardat's method to calculate the synthetic magnitude, where the results are displayed in Figure (3) and (4).
(d) The spectral types of both components are identified as *F*9 and *K*3, A and B respectively.

## 6. ACKNOWLEDGMENTS:

The authors uses the codes of Al-Wardat's method for analyzing binary and multiple stellar systems.

The authors also thank Mounib Eid for providing his comments and suggestions. The authors would also like to thank the anonymous referee for the careful reading of the manuscript and for all suggestions and comments which allowed us to improve both the quality and the clarity of the paper. This work used the SAO/NASA database, Gaia Data (Gaia DR2) and (EDR3), the SIMBAD database, the fourth catalogue of interferometry measurements of binary stars. IPAC data systems.